# Optimal Resource Allocation for OFDM Multiuser Channels

Gerhard Wunder and Thomas Michel

German-Sino Mobile Communications Lab at

Fraunhofer Institute for Telecommunications, Heinrich-Hertz-Institut

Einstein-Ufer 37, D-10587 Berlin

{wunder,michel}@hhi.fhg.de

**Abstract**

In this paper, a unifying framework for orthogonal frequency division multiplexing (OFDM) multiuser resource allocation is presented. The isolated seeming problems of maximizing a weighted sum of rates for a given power budget $\bar{P}$ and minimizing sum power for given rate requirements $\bar{\mathbf{R}}$ can be interpreted jointly in this framework. To this end we embed the problems in a higher dimensional space. Based on these results, we subsequently consider the combined problem of maximizing a weighted sum of rates under given rate requirements $\bar{\mathbf{R}}$ and a fixed power budget $\bar{P}$. This new problem is challenging, since the additional constraints do not allow to use the hitherto existing approaches. Interestingly, the optimal decoding orders turn out to be the ordering of the Lagrangian factors in all problems.

I. INTRODUCTION

Resource allocation has become a crucial challenge in modern digital wireless communication systems, likewise for the downlink as well as for the uplink. Limited system resources such as power, bandwidth and transmit time have to be assigned in an optimal way, where optimality

This work was supported by the German Ministry for Education and Research under Grant FK 01 BU 350 (the *3GET* project). Parts of this paper were presented at Allerton Conference 2005 [1] and ICASSP 2006 [2].

depends very much on the requirements of the subscribers and the operator: The objective is to maximize a benefit (or minimize a cost) function such as throughput (or sum power) under certain constraints, which might be power, delay or other Quality of Service (QoS) constraints imposed by the system and the users' services. For example, a well known objective function is the weighted rate-sum, constrained by a given power budget. Choosing the weights to be the buffer sizes is closely related to the stability of the system [3], [4], but in principle the choice of the weights might be of any kind. On the other hand sum power is an important cost function, and fairness or target rates are common constraints [5]–[8]. These problems seem isolated and the connection between them - if any - remains unclear.

Especially the first optimization problem is considered, when multiuser capacity regions are characterized: For the flat fading case, the delay limited and ergodic capacity regions of the multiple access channel (MAC) have been studied in [9], [10] using the weighted sum rate-approach. The equivalent broadcast channel (BC) was treated in [11], [12]. A general duality between MAC and BC for perfect channel state information was established in [13], allowing to translate results from one multiuser channel to the other.

Since OFDM has been implemented in promising standards such as digital audio/video broadcasting (DAB, DVB), wireless local area networking (WLAN), and it emerged that OFDM is becoming a key technology in future 4G systems, resource allocation for OFDM has been investigated intensively [14]–[18]. Using the weighted-sum rate approach it was shown that frequency division multiple access (FDMA) is optimal to achieve the maximum sum rate for a sum power constraint [14] and for individual power constraints [17]. This is not the case for any weighted rate sum optimization problem, i.e. for other points on the boundary of the capacity region. The sum power minimization problem was studied in [7], leading to a nonconvex formulation. The general problem in finding the minimum sum power is, that the OFDM channel is a *non-degraded* channel: this makes it extremely complicated to determine the optimal decoding order. The problem does not affect the maximization of a weighted rate-sum, since the decoding order can be found exploiting the polymatroid structure of the capacity region.

In this paper, we find the minimum sum power of the OFDM multiuser channel for required rates. We show, that a wide class of problems can be formulated in a unifying framework embedded in a higher dimensional space based on Lagrangian theory. This establishes a connec-tion between the optimization problems. Being aware of the coherence, we can derive efficient algorithms for all of these problems. Furthermore, we can combine the two antagonistic views of minimum rates and weighted rate sum-maximization to a more sophisticated problem, taking into account both necessities similar to [19], but for the frequency selective case. Furthermore, the resulting vector of Lagrangian parameters reveals the optimal decoding (encoding) order.

The remainder of this paper is organized as follows: Section II describes the system model and Section III states the problem formulation. In Sections IV-VI the solutions to the specific problems are derived and the connections among them are depicted. Finally, we conclude with Section VIII.

## II. System Model

In this paper, we use the following notation: Vectors and matrices are written bold face and sets are written in calligraphic letters. The notation $\mathbf{A} = \mathrm{diag}_{k=1}^{K}\{a_k\}$ denotes the $K \times K$ diagonal matrix with the entries $A_{k,k} = a_k$ on its main diagonal. $\mathbb{E}\{\cdot\}$ is the expectation operator and $|.|_1$ the $l_1$-norm. A circularly symmetric Gaussian random variable $Z = X + jY \sim \mathcal{CN}(0, \sigma^2)$ is a random variable with i.i.d. real variables $X, Y \sim \mathcal{N}(0, \sigma^2/2)$ and all logarithms are to the base $e$.

We assume an OFDM system with $K$ subcarriers, a base station and $M$ users. Further, we suppose perfect channel state information at both sides. The set of users will be denoted by $\mathcal{M} = \{1, ..., M\}$ in the following. The channel between the base station and any user $m \in \mathcal{M}$ is modeled as a frequency selective channel with $L_m$ taps. Denoting the $l$th tap of the channel impulse response of user $m$ as $h_m[l]$, the channel on carrier $k$ can be written as

$$h'_{m,k} = \sum_{l=1}^{L_m} h_m[l] e^{-2\pi j(l-1)(k-1)/K}. \tag{1}$$

We apply a cyclic prefix (CP) of length $L_{CP} \geq \max_m L_m$ to ensure orthogonality of the subcarriers and neglect the loss of spectral efficiency due to the CP assuming that it is small.

Now we consider two different systems: the broadcast channel, where the base station sends independent messages to each user and the multiple access channel, where each user transmits independent messages to the base station. Assuming channel reciprocity the system equations

on each subcarrier $k$ can be written as

$$y_{m,k} = h'_{m,k} \sum_{s=1}^{M} x_{s,k} + n \qquad (2)$$

$$y_k = \sum_{m=1}^{M} h'_{m,k} x_{m,k} + n \qquad (3)$$

where (2) is the broadcast channel model and (3) the multiple access channel model. In each case, the signal is scaled by the channel gain and corrupted by circularly symmetric white Gaussian noise $n \sim \mathcal{CN}(0, \sigma_n^2)$. Note, that the receiver thermal noise power is assumed to be equal in both cases. If the channel gains $h_{m,k} = |h'_{m,k}|^2$ are the same for the uplink and the downlink, we call MAC and BC dual. We assume perfect channel state information, that is, the channel is perfectly known at both ends in each fading state. Due to this assumption, the following is not restricted to our fading model but holds for any known channel realization.

## A. The OFDM broadcast and multiple access channel

Let $\mathbf{h} = (h_{1,1}, \ldots, h_{1,K}, h_{2,1}, \ldots, \ldots, h_{M,K})^T$ denote the stacked vector of channel gains and $\mathbf{h}_k = (h_{1,k}, \ldots, h_{M,k})^T$ the vector of channel gains of all users on subcarrier $k$. It is known that with Costa Precoding the entire capacity region of the non-degraded broadcast channel, such as the OFDM BC, can be achieved. Note, that the OFDM BC region can also be achieved with Superposition Coding together with Successive Interference Cancellation [20], since it can be seen as a set of parallel Gaussian degraded broadcast channels. Let $\pi \in \Pi$ be an arbitrary Costa encoding order from the set of all possible encoding orders $\Pi$. Then user $\pi(M)$ is encoded first, followed by user $\pi(M-1)$ and so on. The base station has full non-causal knowledge of the message transmitted to the previously encoded users and can take this knowledge into account when choosing the codeword for the following user. Denote the power of user $m$ on carrier $k$ by $p_{m,k} = \mathbb{E}\{|x_{m,k}|^2\}$, the stacked vector of powers as $\mathbf{p} = (p_{1,1}, \ldots, p_{1,K}, p_{2,1}, \ldots, \ldots, p_{M,K})^T$ and the power vector of user $m$ as $\mathbf{p}_m = (p_{m,1}, \ldots, p_{m,K})^T$. Then the rate of user $\pi(m)$ can be expressed as

$$\tilde{R}_{\pi(m)} = \sum_{k=1}^{K} \log \left( 1 + \frac{h_{\pi(m),k} p_{\pi(m),k}}{\sigma^2 + h_{\pi(m),k} \sum_{n<m} p_{\pi(n),k}} \right). \qquad (4)$$

The capacity region[1] of the OFDM BC under a given sum power constraint $\bar{P}$ is given by

$$\mathcal{C}_{BC}(\mathbf{h}, \bar{P}) \equiv \bigcup_{\substack{\pi \in \Pi \\ |\mathbf{p}|_1 \leq \bar{P}}} \left\{ \mathbf{R} : R_{\pi(m)} = \tilde{R}_{\pi(m)} \,,\, m \in \mathcal{M} \right\} \tag{5}$$

where $\tilde{R}_{\pi(m)}$ is defined in eq. (4). Using the same stacked vector notation $\mathbf{h}$, the capacity region of the dual MAC under the same sum power constraint $\bar{P}$ can be expressed as

$$\mathcal{C}_{MAC}(\mathbf{h}, \bar{P}) \equiv \bigcup_{|\mathbf{p}|_1 = \bar{P}} \left\{ \mathbf{R} : \sum_{m \in \mathcal{S}} R_{\pi(m)} \leq \sum_{k=1}^{K} \log\left(1 + \frac{1}{\sigma^2} \sum_{m \in \mathcal{S}} h_{m,k} p_{m,k}\right) \,,\, \mathcal{S} \subseteq \mathcal{M} \right\}. \tag{6}$$

Assume ideal successive interference cancellation (SIC) and an arbitrary decoding order $\pi \in \Pi$. Then user $\pi(1)$ is decoded first, followed by user $\pi(2)$ and so on. An equivalent characterization of the OFDM MAC, using successive interference cancellation (SIC) is given by

$$\mathcal{C}_{MAC}(\mathbf{h}, \bar{P}) \equiv \bigcup_{\substack{\pi \in \Pi \\ |\mathbf{p}|_1 \leq \bar{P}}} \left\{ \mathbf{R} : R_{\pi(m)} = \tilde{R}'_{\pi(m)} \,,\, m \in \mathcal{M} \right\} \tag{7}$$

where $\tilde{R}'_{\pi(m)}$ is given by

$$\tilde{R}'_{\pi(m)} = \sum_{k=1}^{K} \log\left(1 + \frac{h_{\pi(m),k} p_{\pi(m),k}}{\sigma^2 + h_{\pi(m),k} \sum_{n>m} p_{\pi(n),k}}\right). \tag{8}$$

Indicating the relation to the broadcast channel, this characterization leads straight to the duality of both systems, treated in the next section.

## B. Duality of OFDM BC and OFDM MAC

A popular method to solve problems for the broadcast channel is to solve them for the MAC and to relate them to the BC by uplink-downlink duality. Such duality results between uplink and downlink were reported for quite a few scenarios in recent years [6], [13], [21], [22]. In this work, we exploit uplink-downlink duality as well. Hence, we state the OFDM duality result in the following.

---

[1]In a slight misuse of the term capacity, we denote the *spectral efficiency* [bits/s/Hz] as capacity. Further, the expression *capacity* and *instantaneous capacity* (denoting the capacity of a single fading state of infinite length) are used equivalently throughout this paper unless specified in a differing manner.

*Lemma 1:* Let $\mathcal{C}_{BC}(\mathbf{h}, \bar{P})$ denote the capacity region of an OFDM BC under a sum power constraint $\bar{P}$ and channel realizations $\mathbf{h}$. In analogy, let $\mathcal{C}_{MAC}(\mathbf{h}, \bar{P})$ denote the capacity region of the OFDM MAC with the same parameters $\bar{P}$ and $\mathbf{h}$. Then both regions coincide, i.e.

$$\mathcal{C}_{BC}(\mathbf{h}, \bar{P}) \equiv \mathcal{C}_{MAC}(\mathbf{h}, \bar{P}). \tag{9}$$

*Proof:* It is easy to see that the OFDM BC can be written as a multiple input multiple output (MIMO) BC simply by stacking the vectors of the signals $\mathbf{x}_m = (x_{m,1}, ..., x_{m,K})^T$, $\mathbf{y}_m = (y_{m,1}, ..., y_{m,K})^T$, $\mathbf{n} = (n_1, ..., n_K)^T$ and setting $\mathbf{H}_m = \mathrm{diag}_{k=1}^{K}\{h'_{m,k}\}$. The same holds for the OFDM MAC.

On the other hand we know from the line of work [21]–[23], that duality holds for the instantaneous capacity regions of Gaussian MIMO BC $\mathcal{C}_{BC}(\{\mathbf{H}_m\}_{m=1}^M, \bar{P})$ and Gaussian MAC $\mathcal{C}_{MAC}(\{\mathbf{H}_m^H\}_{m=1}^M, \bar{P})$. Combining the reformulation of the OFDM BC (MAC) in terms of MIMO BC (MAC) with the duality result concludes the proof. ∎

*Remark 1:* Note, that the duality result holds independent of the number of transmit and receive antennas, since any MIMO-OFDM system can be written in block-diagonal form [24]. Only the conjugate transposition has to be taken into account in case of more than one antenna.

It is known from [13], [21] that both regions are achieved by circular symmetric complex i.i.d. Gaussian symbols $x_{m,k} \sim \mathcal{CN}(0, p_{m,k})$ and if a set of rates $\mathbf{R} \in \mathcal{C}_{BC}(\mathbf{h}, \bar{P})$ is achieved by a power allocation $\mathbf{p}^{BC} = (p^{BC}_{\pi(1),1}, \ldots, p^{BC}_{\pi(M),K})^T$, the same set of rates is achieved in the dual MAC by the power allocation

$$\begin{aligned}
p^{MAC}_{\pi(M),k} &= p^{BC}_{\pi(M),k} \frac{\sigma^2}{\sigma^2 + h_{\pi(M),k} \sum_{n=1}^{M-1} p^{BC}_{\pi(n),k}} \\
p^{MAC}_{\pi(M-1),k} &= p^{BC}_{\pi(M-1),k} \frac{\sigma^2 + h_{\pi(M),k} p^{BC}_{\pi(M),k}}{\sigma^2 + h^2_{\pi(M-1),k} \sum_{n=1}^{M-2} p^{BC}_{\pi(n),k}} \\
&\ldots \\
p^{MAC}_{\pi(1),k} &= p^{BC}_{\pi(M),k} \frac{\sigma^2 + \sum_{n=2}^{M} h_{\pi(n),k} p^{BC}_{\pi(n),k}}{\sigma^2}
\end{aligned} \tag{10}$$

Inverse relations - transforming the power allocations from MAC to BC - can be found in [13], Eq. (12).

Note, that due to the established duality, it does not matter whether a problem is solved in the OFDM MAC or the OFDM BC, since the solution can be transfered from one to the other.

Hence, we can choose the easiest formulation among MAC and BC for each problem. In the remainder of the paper, the subscripts $\cdot_{MAC}$ and $\cdot_{BC}$ are omitted and all variables refer to the multiple access channel unless specified in a differing manner.

### III. PROBLEM FORMULATION AND GLOBAL LAGRANGIAN PERSPECTIVE

In this paper, we want to solve the following three problems and relate them one to each other:

*Problem 1:* **[Weighted Sum Rates Problem]**

$$\begin{aligned} \text{maximize} \quad & \boldsymbol{\mu}^T \mathbf{R} \\ \text{subject to} \quad & \mathbf{R} \in \mathcal{C}(\mathbf{h}, \bar{P}) \end{aligned} \quad (11)$$

This problem stems from an operators perspective. The objective is to maximize a weighted sum of rates subject to a sum power constraint $\bar{P}$. The weights represent relative priorities. The most common is to take the buffer length weighted rate-sum, which is related to the stability region [25]. In principle, the weights $\boldsymbol{\mu}$ can have any other interpretation. The approach yields a parameterization of the boundary of the capacity region.

*Problem 2:* **[Minimum Sum Power Problem]**

$$\begin{aligned} \text{minimize} \quad & P \\ \text{subject to} \quad & R_i \geq \bar{R}_i \quad \forall i \in \mathcal{S}, \mathcal{S} \subseteq \mathcal{M} \\ & \mathbf{R} \in \mathcal{C}(\mathbf{h}, P) \end{aligned} \quad (12)$$

Minimizing the sum power, which is necessary to achieve certain rate requirements $\bar{\mathbf{R}} = (\bar{R}_1, ..., \bar{R}_M)^T$, is the objective of the second problem. Since the system is power limited but not interference limited, any set of finite rates $\bar{\mathbf{R}}$ can be achieved. A crucial point is to determine the optimal decoding order.

*Problem 3:* **[Minimum Rates Problem]**

$$\begin{aligned} \text{maximize} \quad & \boldsymbol{\mu}^T \mathbf{R} \\ \text{subject to} \quad & R_i \geq \bar{R}_i \quad \forall i \in \mathcal{S}, \mathcal{S} \subseteq \mathcal{M} \\ & \mathbf{R} \in \mathcal{C}(\mathbf{h}, \bar{P}) \end{aligned} \quad (13)$$

This problem is a combination of the two previous ones. It is more realistic and at the same time more sophisticated. The first condition reflects the minimum rate requirements $\bar{\mathbf{R}}$ and the second stems from the limited power budget $\bar{P}$. Once again, the optimal decoding order is unknown. Additionally, the question of feasibility arises, since not every set of required rates $\bar{\mathbf{R}}$ may be supportable with the given power budget $\bar{P}$.

In the remainder of this paper, we will solve the three problems - each in frameworks - and show the relation between them.

## A. A global Lagrangian perspective

We begin by deriving a unique framework for all problems in a higher dimensional space. By the concavity of the $log$-function it is easy to prove the following lemma:

*Lemma 2:* The set $\mathcal{G}(\mathbf{h}) = \{(\mathbf{R}, P) : \mathbf{R} \in \mathcal{C}(\mathbf{h}, P)\}$ is a convex set.

Obviously, the duality of MAC and BC holds for the enhanced set $\mathcal{G}(\mathbf{h})$, too. By Lemma 2 there are $\lambda \in \mathbb{R}_+$ and $\boldsymbol{\mu} \in \mathbb{R}_+^M$ such that Problems 1-3 are equivalent to solving the following problem:

$$\max \ \boldsymbol{\mu}^T \mathbf{R} - \lambda P \ \text{subj. to} \ (\mathbf{R}, P) \in \mathcal{G}(\mathbf{h}) \tag{14}$$

Note, that this reformulation is more than a formal step. So each point on the boundary of the set $\mathcal{G}(\mathbf{h})$ can be parameterized as the solution to the above problem for a specific choice of the parameters $(\boldsymbol{\mu}, \lambda)$. Besides the Lagrangian interpretation as the direction of steepest descent, the vector $(\boldsymbol{\mu}, \lambda)$ can be interpreted as the normal vector of a tangent hyperplane to the solution. Further, the stated problem is convex, since the function in (14) is affine and the set $\mathcal{G}(\mathbf{h})$ is a convex set.

Each of the Problems 1-3 can be interpreted in the introduced framework: In Problem 1 the Lagrangian vector $\boldsymbol{\mu}$ is known. The factor $\lambda$ is unknown a priori and can be seen as the *power price*. In contrast, in Problem 2 the vector $\boldsymbol{\mu}$ constitutes the unknown Lagrangian vector of the rate requirements. It turns out, that the $\mu_m$ have the additional interpretation as individual water-filling levels for rate water-filling. The factor $\lambda = 1$ is fixed a priori, where it could equivalently be fixed to any other positive real number; it results only in a scaling of the other Lagrangian parameters. This is equivalent to the normalization of the vector $\boldsymbol{\mu}$ in Problem 1, since the scaling of the normal vector $(\boldsymbol{\mu}, \lambda)$ is irrelevant. Thus, the number of degrees of freedom is one less than the number of Lagrangian multipliers.

The relation to the last Problem 3 will be pointed out in the remainder of this paper. All problems have in common, that we search for parts of the Lagrangian normal vector to the supporting hyperplane and parts of the optimal supported point on the boundary of $\mathcal{G}(\mathbf{h})$.

## IV. MAXIMUM WEIGHTED SUM-RATE FOR GIVEN POWER BUDGET

Although a well studied topic, we will derive an uplink solution to Problem 1 briefly for two reasons: First, the methods are of significant relevance for other systems such as multiuser MIMO, and further the derivation allows the characterization of orthogonal transmission modes, which are given in Lemma 3.

The key to the reformulation as a convex problem is the polymatroid structure of the capacity region $\mathcal{C}(\mathbf{h}, \hat{\mathbf{p}})$ for a fixed power allocation $\hat{\mathbf{p}}$. The solution to the maximization of a linear function $g(\mathbf{R}) = \boldsymbol{\mu}^T \mathbf{R}$ over a polymatroid is known to be the vertex corresponding to the decreasing ordering $\pi$ of the factors $\mu_m$ of $g(\mathbf{R})$ [26]. Taking the union over all power allocations in (7) does not affect the decoding order and hence the optimal decoding order is known to be $\pi$. This leads to a convex formulation, whose Lagrangian is given by

$$\mathcal{L}(\mathbf{p}, \boldsymbol{\mu}, \tilde{\lambda}, \tilde{\boldsymbol{\phi}}) = \sum_{m=1}^{M} \mu_{\pi(m)} \log \left( 1 + \frac{p_{\pi(m),k} h_{\pi(m),k}}{\sigma^2 + \sum_{n>m} p_{\pi(n),k} h_{\pi(n),k}} \right) + \tilde{\lambda}(|\mathbf{p}|_1 - \bar{P}) + \sum_{m=1}^{M} \sum_{k=1}^{K} \tilde{\phi}_{m,k} p_{m,k}, \tag{15}$$

where $\tilde{\phi}_{m,k}$ is the Lagrangian multiplier due to the nonnegativity of the power $p_{m,k}$. Note, that $\mathcal{L}(\mathbf{p}, \boldsymbol{\mu}, \tilde{\lambda}, \tilde{\boldsymbol{\phi}})$ is a convex function in $\mathbf{p}$ since due to the definition of the ordering $\pi$ we have $\mu_{\pi(m)} \geq \mu_{\pi(m+1)}$. Hence, the Karush-Kuhn-Tucker (KKT) conditions are necessary and sufficient for optimality and any standard convex optimization method can be applied to solve the problem. $\tilde{\lambda}$ is the Lagrangian factor belonging to the power constraint and $\tilde{\phi}_{m,k}$ is the Lagrangian multiplier due to non-negativity of $p_{m,k}$. By setting $c_m = \mu_{\pi(m)} - \mu_{\pi(m-1)}$ for all $m \neq 1$ and $c_1 = \mu_{\pi(1)}$,

the KKT conditions can be written as:

$$\sum_{s=1}^{m} \frac{c_s h_{\pi(m),k}}{\sigma^2 + \sum_{n=s}^{M} p_{\pi(n),k} h_{\pi(n),k}} - \tilde{\lambda} + \tilde{\phi}_{m,k} = 0 \tag{16}$$

$$\sum_{k=1}^{K} p_{m,k} - \bar{P} \leq 0 \tag{17}$$

$$\tilde{\lambda} \left( \sum_{k=1}^{K} p_{m,k} - \bar{P} \right) = 0 \tag{18}$$

$$\tilde{\lambda} \geq 0, \quad p_{m,k} \geq 0, \quad \tilde{\phi}_{m,k} \geq 0, \quad p_{m,k} \tilde{\phi}_{m,k} = 0 \tag{19}$$

for all $m$ and $k$. The KKT conditions will be used to derive Lemma 3.

## A. Algorithmic Solution

An alternative algorithm much more efficient than standard convex optimization was presented in unpublished work [20]: Tse derived an algorithm for solving this problem for parallel Gaussian channels directly in the downlink. Since the OFDM BC can be interpreted as a set of parallel Gaussian channels (as long as the CP is long enough and orthogonality can be guaranteed) the algorithm can be carried over to the OFDM BC case. He introduces the elegant notion of *marginal utility functions* to characterize the revenue of each user to the objective function:

$$u_m^{(k)}(z) = \frac{\mu_m}{\left(\frac{\sigma^2}{h_{m,k}} + z\right)} - \tilde{\lambda} \tag{20}$$

The set of equations characterizing the solution is given by

$$R_m = \int_0^{\infty} \sum_{k: u_m^{(k)}(z) = \left[\max_i u_i^{(k)}(z)\right]^+} \frac{1}{\left(\frac{\sigma^2}{h_{m,k}} + z\right)} dz \tag{21}$$

$$P(\tilde{\lambda}) = \sum_{k=1}^{K} \left[ \max_m \left( \frac{\mu_m}{\tilde{\lambda}} - \frac{\sigma^2}{h_{m,k}} \right) \right]^+ \tag{22}$$

where $\tilde{\lambda}$ is the Lagrangian parameter of the sum power constraint and $[a]^+ := \max(0, a)$. For a detailed study the reader is referred to [9], [20]. The solution can be found by solving equation (22) for the Lagrangian multiplier $\tilde{\lambda}$. This can be done by simple bisection, since the RHS of (22) is monotone in $\tilde{\lambda}$ and equation (22) yields a simple expression for the sum power constraint $\bar{P}$ as a function of the Lagrangian parameter $\tilde{\lambda}$. Subsequently, the relevant intersections have to be calculated, which is not a combinatorial problem, since the ordering of channel coefficients defines the ordering of intersections. Hence, Algorithm 1 presented below is more efficient than standard convex optimization and than *fixed level-water-filling*, which will be introduced shortly in Section V.

---
**Algorithm 1** Weighted rate sum maximization

   **(1)** solve (22) for Lagrangian factor $\lambda$

  **for** $k = 1$ to $K$ **do**

     **(2)** order users according to their channel gains

     **(3)** determine intersections of marginal utility functions (20)

  **end for**

   **(4)** calculate resulting rates (21)

---

## B. Special transmission modes

It is desirable to know, in which cases orthogonal signaling is optimal due to the easy signal processing in this case. The arising problem is to find conditions for the optimality of FDMA and the single user range. The Karush-Kuhn-Tucker (KKT) conditions (16) - (19) from the convex formulation in the uplink scenario allow such a proposition:

*Lemma 3:* Assume that the user weights are ordered, i.e. $\mu_1 \geq ... \geq \mu_M \geq 0$. Let the index of the user transmitting exclusively on carrier $k$ be $m_k$. Then exclusive subcarrier assignment (FDMA), i.e. allocating the power $\bar{P}$ to user $m_k$ on carrier $k$ for all $k = 1, ..., K$ is optimal if and only if

$$\max_k \frac{\mu_{m_k} h_{m_k,k}}{\sigma^2 + p_{m_k,k} h_{m_k,k}} \geq \max_{\substack{m \neq m_k \\ k}} \left\{ \max(\mu_m - \mu_{m_k}, 0) \frac{h_{m,k}}{\sigma^2} + \min(\mu_{m_k}, \mu_m) \frac{\mu_m h_{m,k}}{\sigma^2 + p_{m_k,k} h_{m_k,k}} \right\}$$
(23)

*Corollary 1:* Single user transmission, i.e. allocating the power $\bar{P}$ exclusively to user $m_{su}$ is optimal if and only if

$$\max_k \frac{\mu_{m_{su}} h_{m_{su},k}}{\sigma^2 + p_{m_{su},k} h_{m_{su},k}} \geq \max_{\substack{m \neq m_{su} \\ k}} \left\{ \max(\mu_m - \mu_{m_{su}}, 0) \frac{h_{m,k}}{\sigma^2} + \min(\mu_{m_{su}}, \mu_m) \frac{\mu_m h_{m,k}}{\sigma^2 + p_{m_{su},k} h_{m_{su},k}} \right\}$$
(24)

*Proof:* From the first KKT condition (16) follows, that $\tilde{\phi}_{m,k} > 0$ for users $m \neq m_k$ and $\tilde{\phi}_{m_k,k} = 0$ on carrier $k$, since $p_{m,k} = 0$ for all $m \neq m_k$. Hence, the KKT-condition (16) for user $m_k$, i.e. the user transmitting exclusively on the $k$th subcarrier is given by:

$$\tilde{\lambda} = \frac{\mu_{m_k} h_{m_k,k}}{\sigma^2 + p_{m_k,k} h_{m_k,k}}$$

For users with index $m < m_k$ we can derive analogously

$$\tilde{\lambda} = \frac{\mu_m h_{m,k}}{\sigma^2 + p_{m_k,k} h_{m_k,k}} + \tilde{\phi}_{m,k}.$$

In case $m > m_k$, the KKT-condition results in

$$\tilde{\lambda} = \frac{\mu_{m_k} h_{m,k}}{\sigma^2 + p_{m_k,k} h_{m_k,k}} + (\mu_m - \mu_{m_k})\frac{h_{m,k}}{\sigma^2} + \tilde{\phi}_{m,k}$$

since the interference term in the denominator vanishes for addends $m > m_k$. Thus, we get the necessary and sufficient conditions

$$\tilde{\lambda} = \max_k \frac{\mu_{m_k} h_{m_k,k}}{\sigma^2 + p_{m_k,k} h_{m_k,k}} \geq \max_{\substack{m < m_k \\ k}} \frac{\mu_m h_{m,k}}{\sigma^2 + p_{m_k,k} h_{m_k,k}} \qquad (25)$$

$$\tilde{\lambda} = \max_k \frac{\mu_{m_k} h_{m_k,k}}{\sigma^2 + p_{m_k,k} h_{m_k,k}} \geq \max_{\substack{m > m_k \\ k}} \frac{\mu_{m_k} h_{m,k}}{\sigma^2 + p_{m_k,k} h_{m_k,k}} \quad + (\mu_m - \mu_{m_k}) h_{m,k}/\sigma^2. \qquad (26)$$

Combining equations (25) and (26) leads to Lemma 3. The condition for the single user optimality range follows immediately from setting $m_1 = ... = m_K$. ∎

*Remark 2:* Note that in the case of sum rate, the condition in Lemma 3 is always fulfilled. Hence, FDMA achieved the full sum rate in OFDM BC and OFDM MAC systems.

At first glance the condition in Lemma 3 seems complicated, since the powers themselves occur in the expression. Nevertheless, a simple water-filling algorithm using the weighted channel gains $\mu_m h_{m,k}$ as decision criterion yields the optimal allocation *in case of FDMA optimality*, which is for example guaranteed in the low SNR regime. Once done, the condition can be checked. Even if FDMA is not optimal, the resource allocation algorithm yields excellent results as shown in [27].

## V. MINIMUM SUM POWER FOR GIVEN RATE REQUIREMENTS

In this section, we change perspectives and derive the solution to the minimum sum power problem for given rate requirements $\bar{\mathbf{R}}$. The main problem is, that the encoding order is not clear, since the exploitation of the polymatroid structure in section IV can not be transfered to the sum power minimization problem. In order to simplify the problem, we will choose the uplink perspective one again, such that the unknown encoding order is equivalent to the reverse SIC order.

A recent approach [7] proposed to divide the problem in two subproblems: first, finding the optimal decoding order and second, solve the problem for a fixed decoding order. We show, that the split into two subproblems falls short of the structure of the problem, and both problems can be solved within one single framework.

Problem 2 can be written as

$$\min_{\pi \in \Pi} \ |\mathbf{p}|_1$$
$$\text{subj. to} \ \log\left(1 + \frac{p_{\pi(m),k} h_{\pi(m),k}}{\sigma^2 + \sum_{n>m} p_{\pi(n),k} h_{\pi_s(n),k}}\right) \geq \bar{R}_{\pi(m)} \quad (27)$$
$$p_{m,k} \geq 0$$

where all powers $p_{m,k}$ are nonnegative. Unfortunately this formulation is not convex, since the constraints are not convex in $\mathbf{p}$.

First, we reformulate the problem in terms of the rate $R_{m,k}$. To this end, solving

$$R_{\pi(m),k} = \log\left(1 + \frac{p_{\pi(m),k} h_{\pi(m),k}}{\sigma^2 + \sum_{n>m} p_{\pi(n),k} h_{\pi_s(n),k}}\right) \quad (28)$$

for $p_{\pi(m),k}$ yields after some manipulations

$$p_{\pi(m),k} = \frac{\sigma^2}{h_{\pi(m),k}} \left(\exp\left(R_{\pi(m),k}\right) - 1\right) \exp\left(\sum_{n>m} R_{\pi(n),k}\right). \quad (29)$$

Now with the substitution (29), the problem can be written as

$$\min_{\pi \in \Pi} \ \sum_{m=1}^{M} \sum_{k=1}^{K} \frac{\left(\exp\left(R_{\pi(m),k}\right) - 1\right)}{h_{\pi(m),k}/\sigma^2} \exp\left(\sum_{n>m} R_{\pi(n),k}\right)$$
$$\text{subj. to} \ \sum_{k=1}^{K} R_{m,k} \geq \bar{R}_m \quad (30)$$
$$R_{m,k} \geq 0.$$

Note, that the constraints are linear now, but the objective function in (30) is not convex any more. To fix the decoding order $\pi$ a priori does not help to solve the problem, since the objective function is *not convex independent of the decoding order*. Surprisingly, this holds even for the optimal decoding order.

To proceed further and to circumvent this problem, we need the following lemma:

*Lemma 4:* Let $\mathcal{C}(\mathbf{h}, \bar{P})$ denote the capacity region of an OFDM MAC under a total sum power constraint $\bar{P}$. The OFDM MAC capacity region $\mathcal{C}(\mathbf{h}, \bar{P})$ can be expressed as

$$\mathcal{C}(\mathbf{h}, \bar{P}) \equiv \bigcup_{\sum_{k=1}^{K} P^{(k)} = \bar{P}} \sum_{k=1}^{K} \mathcal{C}^{\pi_k}(\mathbf{h}_k, P^{(k)}) \tag{31}$$

where $\mathbf{h}_k = (h_{1,k}, \ldots, h_{M,k})^T$ and $\pi_k \in \Pi$ are decoding orders such that

$$h_{\pi_k(1),k} \geq \ldots \geq h_{\pi_k(M),k}, \quad \forall k, \tag{32}$$

and

$$\mathcal{C}^{\pi_k}(\mathbf{h}_k, P) \equiv \bigcup_{\sum_m^M p_m = P} \left\{ \mathbf{R} : R_{\pi_k(m)} \leq \log\left(1 + \frac{p_{\pi_k(m)} h_{\pi_k(m)}}{\sigma^2 + \sum_{n > m} p_{\pi_k(n)} h_{\pi_k(n)}}\right) \forall m \in \mathcal{M} \right\} \tag{33}$$

is the achievable rate region for a given decoding order $\pi_k$.

*Proof:* We know from [9], [20] that

$$\mathcal{C}(\mathbf{h}, \bar{P}) \equiv \bigcup_{\sum_{k=1}^{K} P^{(k)} = \bar{P}} \sum_{k=1}^{K} \mathcal{C}(\mathbf{h}_k, P^{(k)}). \tag{34}$$

Further, the optimal decoding order for a scalar Gaussian MAC is given by $\pi_{opt}$ such that

$$h_{\pi_{opt}(1),k} \geq \ldots \geq h_{\pi_{opt}(M),k} \quad \forall k. \tag{35}$$

With this decoding order, every point of the capacity region $\mathcal{C}(\mathbf{h}_k, P^{(k)})$ of the degraded channel can be achieved; this is well known, see e.g. [9]. Hence, in the special case of a scalar MAC (i.e. on each subcarrier), the capacity region coincides with the region $\mathcal{C}^{\pi_{opt}}(\mathbf{h}_k, P^{(k)})$, i.e.

$$\mathcal{C}(\mathbf{h}_k, P^{(k)}) \equiv \mathcal{C}^{\pi_{opt}}(\mathbf{h}_k, P^{(k)}). \tag{36}$$

Substituting (36) in (35) completes the proof. ∎

Lemma 4 states, that the OFDM MAC (the same holds for the OFDM BC, see [20]) can be decomposed to a union over the sum of scalar MACs, which are MAC channels where the whole region can be achieved with a single decoding order. Each scalar MAC represents a specific subcarrier. This representation will turn out to be the key to an efficient solution to the sum power minimization problem, since it circumvents the problem of finding the globally optimal decoding order. Motivated by Lemma 4, we introduce an individual decoding order on each subcarrier denoted by $\pi_k$, such that

$$h_{\pi_k(1),k} \geq \ldots \geq h_{\pi_k(M),k}, \quad \forall k, \tag{37}$$

and define coefficients

$$c_{m,k} = \sigma^2 \left( h^{-1}_{\pi_k(m),k} - h^{-1}_{\pi_k(m-1),k} \right) \quad \forall m \in \mathcal{M} \setminus \{1\} \tag{38}$$

and

$$c_{1,k} = h^{-1}_{\pi_k(1),k} \tag{39}$$

which are non-negative due to the definition of the permutations $\pi_k$. Reordering the addends and observing that the remaining term $-h^{-1}_{\pi_k(M),k}$ does not alter the solution we arrive at

$$\begin{aligned}
\min \quad & \sum_{m=1}^{M} \sum_{k=1}^{K} c_{m,k} \exp\left( \sum_{n \geq m} R_{\pi_k(n),k} \right) \\
\text{subj. to} \quad & \sum_{k=1}^{K} R_{m,k} \geq \bar{R}_m \\
& R_{m,k} \geq 0.
\end{aligned} \tag{40}$$

Note, that the objective function in (40) is *log-convex*, since it is a nonnegative sum of log-convex functions. The constraints are linear and the function is defined on a convex set, hence the problem can be solved by standard convex optimization methods.

Let $\mathcal{L}(\mathbf{R}, \tilde{\boldsymbol{\mu}}, \tilde{\boldsymbol{\phi}})$ be the Lagrangian given by

$$\mathcal{L}(\mathbf{R}, \tilde{\boldsymbol{\mu}}, \tilde{\boldsymbol{\phi}}) = \sum_{m=1}^{M} \sum_{k=1}^{K} c_{m,k} \exp\left( \sum_{n \geq m} R_{\pi_k(n),k} \right) - \sum_{m=1}^{M} \tilde{\mu}_m \left( \sum_{k=1}^{K} R_{m,k} - \bar{R}_m \right) - \sum_{m=1}^{M} \sum_{k=1}^{K} \tilde{\phi}_{m,k} R_{m,k} \tag{41}$$

where $\tilde{\mu}_m$ and $\tilde{\phi}_{m,k}$ are non-negative Lagrangian multipliers corresponding to the individual rate constraints and the non negativity of the rates. The Karush-Kuhn-Tucker (KKT) conditions are necessary and sufficient for the optimality of a given set of rates and given by

$$\sum_{s=1}^{m} c_{s,k} \exp\left( \sum_{n \geq s} R_{\pi_k(n),k} \right) - \tilde{\phi}_{m,k} = \tilde{\mu}_m \tag{42}$$

$$\sum_{k=1}^{K} R_{m,k} - \bar{R}_m \geq 0 \tag{43}$$

$$\tilde{\mu}_m \left( \sum_{k=1}^{K} R_{m,k} - \bar{R}_m \right) = 0 \tag{44}$$

$$\tilde{\mu}_m \geq 0, \quad \tilde{\phi}_{m,k} \geq 0, \quad R_{m,k} \geq 0, \quad \tilde{\phi}_{m,k} R_{m,k} = 0 \tag{45}$$

for all $m$ and $k$. The KKT-conditions offer the possibility to design an appealing algorithm based on water-filling, which is computationally much less demanding than convex optimization.

## A. Optimal decoding order

Up to now, the important question of the optimal decoding order remains open: once the problem is solved and a solution to the set of equations (42)-(45) is found, i.e. the optimal power allocation is found, the optimal decoding order can be identified:

One can simply go through the set of subcarriers and check the optimal power allocation on each carrier. For each carrier, a nonempty set of optimal decoding orders $\Pi_{opt}^{(k)}$ exists. These are given by all $\pi_k \in \Pi_{opt}^{(k)}$, such that $m < n$ implies

$$h_{\pi_k(m),k} \geq h_{\pi_k(n),k}, \quad m,n \in \mathcal{S}_k$$

where $\mathcal{S}_k \equiv \{m : p_{m,k} > 0\}$.

In other words, $\Pi_{opt}^{(k)}$ is the set of decoding orders, which do not change the order among all transmitting users. The set of optimal decoding orders $\Pi_{opt}$ is the intersection of all these sets

$$\Pi_{opt} = \bigcap_{k \in \mathcal{K}} \Pi_{opt}^{(k)}. \tag{46}$$

In most cases, the set $\Pi_{opt}$ reduces to one single decoding order $\pi_{opt}$. Since this is a rather heuristic but intuitive explanation, we will make it more rigorous: The ordering of the Lagrangian factors $\tilde{\mu}_m$ yields the optimal decoding order $\pi_{opt}$ such that

$$\tilde{\mu}_{\pi_{opt}(1)} \geq ... \geq \tilde{\mu}_{\pi_{opt}(M)}. \tag{47}$$

This follows directly from the fact, that the capacity region of the optimal power allocation is a polymatroid. Now we have the following interpretation: For each point $\mathbf{R}$ on the boundary of $\mathcal{C}(\mathbf{h}, \bar{P})$, there exists a *global* decoding order $\pi_{opt}$ such that, applying this decoding order on *every* subcarrier, $\mathbf{R}$ is achieved. On the other hand Lemma 4 reveals, that the capacity region is the union over individual MAC regions, whose optimal decoding orders are known. Thus, there exists at least one set of *individual* decoding orders $\{\pi_k\}_{k=1}^K$, such that, applying decoding order $\pi_k$ on subcarrier $k$, the same point is achieved. Applying subcarrier-wise individual decoding orders does not increase the achievable rate region due to Lemma 4. Hence, the entire capacity region can be achieved by a set of decoding orders $\{\pi_k\}_{k=1}^K$ applied to each subcarrier individually. Note, that the assumption of individual decoding orders is made implicitly in [20] by introducing marginal utility functions. The decoding order on subcarrier $k$ corresponds to the ordering of the sets $\mathcal{A}_{n,k} \equiv \{z \in [0,\infty) : u_{n,k}(z) = \max_{i \in \mathcal{M}} u_{i,k}(z)\}$.

## B. Algorithmic solution

To derive an efficient algorithm, we define coefficients $n_{m,k}$ as

$$n_{m,k} = \log \left\{ \exp\left( \sum_{n>m} R_{\pi_k(n),k} \right) \left[ \frac{\sigma^2}{h_{\pi_k(m),k}} \right. \right.$$
$$\left. \left. + \sum_{j=1}^{m-1} \frac{\sigma^2}{h_{\pi_k(j),k}} \left( \exp\left( R_{\pi_k(j),k} \right) - 1 \right) \exp\left( \sum_{n=j+1}^{m-1} R_{\pi_k(n),k} \right) \right] \right\} \quad (48)$$

which can be seen as efficient noise levels. Then the Karush-Kuhn-Tucker (KKT) condition in eq. (42) can be written in a form allowing water-filling:

$$R_{\pi_k(m),k} = \left[ \log(\tilde{\mu}_m) - n_{m,k} \right]^+ \quad \forall m, k \quad (49)$$

Note, that additionally to the interpretation of $\tilde{\mu}_m$ as part of the normal vector of the tangent hyperplane, the term $\log(\tilde{\mu}_m)$ is the *water-filling level* of the rate water-filling.

This motivates Algorithm 2, which shows excellent convergence behavior due to the log-convexity and the semi-analytic stepwise solution through water-filling.

---

**Algorithm 2** Iterative "multi-level rate water-filling"

set $R_{m,k} = 0 \quad \forall m \in \mathcal{M}, \; k = 1, ..., K$

**while** desired accuracy is not reached **do**

  **for** $m = 1$ to $M$ **do**

    **(1)** compute the coefficients $n_{m,k}$ (48) for user $m$

    **(2)** do water-filling with respect to the rates $R_{m,k}$ for user $m$ as in equation (49)

  **end for**

**end while**

---

*Theorem 1:* Let $\hat{\mathbf{R}} = (\hat{R}_{1,1}, ..., \hat{R}_{M,K})^T$ be a vector of rates such that equations (42)-(45) are fulfilled. Algorithm 2 converges to $\hat{\mathbf{R}}$.

*Proof:* We define the set of rates

$$\mathcal{R} = \{R_{m,k} : R_{m,k} \leq \max_m \bar{R}_m \quad m \in \mathcal{M}\}. \quad (50)$$

Since all rate requirements are finite per definition, $\mathcal{R}$ is a compact convex set. The sum power as a function of the actual rates $P(\mathbf{R}), \mathbf{R} \in \mathcal{R}$ is continuously differentiable on $\mathcal{R}$ and bounded from below. The constraints are separable. In each step $n$ of Algorithm 2, the unique water-filling solution with respect to one specific user $m$ is found by fulfilling the individual optimality conditions of this partial problem. So we know that

$$P(\mathbf{R})^{(n+1)} \leq P(\mathbf{R})^{(n)} \qquad (51)$$

where $P(\mathbf{R})^{(n)}$ is the sequence of sum powers. Hence, $P(\mathbf{R})^{(n)}$ is monotonically decreasing and bounded from below. Taking subsequences yields a convergent sequence, which has a limiting point $\hat{\mathbf{R}}$. It is shown in [28], that due to the uniqueness of the solution to each partial step any limiting point is a stationary point. Thus, by the convexity of the problem, each local optimum is a global optimum and hence we conclude, that $\hat{P}(\hat{\mathbf{R}})$ is the minimum sum power. Furthermore, if all channel gains on each subcarrier are distinct, which occurs with probability 1, also the sequence of rates converges to a unique solution. ∎

The convergence behavior is illustrated in Figure 1. The decoding orders $2 \to 3 \to 4 \to 1$ and $2 \to 3 \to 1 \to 4$ turn out to be optimal: the users 1 and 4 do not interfere on any subcarrier. It can be seen very well, that the algorithm from [7] shows an unsteady convergence behavior due to nonconvexity.

## VI. MINIMUM RATES AND GIVEN POWER BUDGET

With the insights of the preceding sections, we turn to Problem 3. Although the problem is very similar to the case studied in [19], the OFDM case is more complicated, since there is no analytical expression for the minimum sum power needed to fulfill the rate requirements. Although not obvious at first glance, Problem 3 can also be interpreted in terms of (14) and thus is a convex problem, which will be shown in this section. Additionally - and indicating the difference to the two previously considered problems - the question of *feasibility* arises. Not all rates might be supportable with a limited power budget $\bar{P}$. Hence, we will have to check feasibility. Another difficulty is, that the decoding order is not obvious. Although the weights $\mu$ are known, they do not determine the optimal decoding order, since they constitute only a part of the tangent hyperplane normal vector.

### A. Convexity and Feasibility

First, we throw a glance on the questions of convexity and feasibility. To this end, we denote the feasible set of Problem 3 by

$$\mathcal{R}_f(\mathbf{h}, \bar{P}) = \{\mathbf{R} : \mathbf{R} \in \mathcal{C}(\mathbf{h}, \bar{P}), R_i \geq \bar{R}_i \ i \in \mathcal{S}\} \qquad (52)$$

where $\mathcal{S}$ is the set of users with rate constraints.

*Lemma 5:* The feasible set $\mathcal{R}_f(\mathbf{h}, \bar{P})$ of Problem 3 in (52) is a convex set. Further, Problem 3 in (11) is feasible, if and only if

$$P_{min} \leq \bar{P},$$

where $\bar{P}$ is the power budget and $P_{min}$ is the solution to Problem 2 with the same given rate requirements $\bar{\mathbf{R}}$ as in Problem 3.

*Proof:* The set $\mathcal{C}(\mathbf{h}, \bar{P})$ is a convex set per definition for all $\mathbf{h}$ and $\bar{P}$, since we can perform time sharing. The sets $\mathcal{R}_f^i = \{\mathbf{R} : R_i \geq \bar{R}_i \ i \in \mathcal{S}\}$ are half spaces bounded by a hyperplane and hence convex sets. Then

$$\mathcal{R}_f(\mathbf{h}, \bar{P}) = \mathcal{C}(\mathbf{h}, \bar{P}) \bigcap \left( \cap_{i \in \mathcal{S}} \mathcal{R}_f^i \right) \tag{53}$$

is the intersection of convex sets and thus a convex set.

A necessary and sufficient condition for feasibility is the nonemptyness of the feasible set. Assume a nonempty feasible set, e.g. $\bar{\mathbf{R}} \in \mathcal{C}(\mathbf{h}, \bar{P})$ and $\bar{P} < P_{min}$. Per definition the rate vector $\bar{\mathbf{R}}$ lies on the boundary of $\mathcal{C}(\mathbf{h}, P_{min})$. Since $\bar{P} < P_{min}$, the region $\mathcal{C}(\mathbf{h}, \bar{P})$ is strictly smaller than $\mathcal{C}(\mathbf{h}, P_{min})$. This leads to a contradiction, since $\bar{R} \in \mathcal{C}(\mathbf{h}, \bar{P})$ and $\bar{R}$ lies on the boundary of a strictly larger region. ∎

This result provides a simple method to answer the question of feasibility. One just has to solve the sum power minimization problem in (13) with Algorithm 2 for the required rates $\bar{\mathbf{R}}$.

The Lagrangian of the problem is given by:

$$\begin{aligned}
\mathcal{L}(\mathbf{R}, \boldsymbol{\mu}, \tilde{\boldsymbol{\mu}}, \tilde{\lambda}, \tilde{\boldsymbol{\phi}}) &= \sum_{m=1}^{M} \mu_{\pi(m)} \sum_{k=1}^{K} R_{\pi(m),k} \\
&+ \sum_{m=1}^{M} \tilde{\mu}_{\pi(m)} \left( \sum_{k=1}^{K} R_{\pi(m),k} - \bar{R}_{\pi(m)} \right) \\
&+ \tilde{\lambda}(|\mathbf{p}|_1 - \bar{P}) + \sum_{m=1}^{M} \sum_{k=1}^{K} \tilde{\phi}_{m,k} p_{m,k}
\end{aligned} \tag{54}$$

where we substituted the rates:

$$R_{\pi(m),k} = \log \left( 1 + \frac{p_{\pi(m),k} h_{\pi(m),k}}{\sigma^2 + \sum_{n>m} p_{\pi(n),k} h_{\pi(n),k}} \right).$$

## B. Algorithmic solution

In this section, we derive two algorithms for Problem 3. Each is derived from one of the two concurring perspectives. The basic idea of Algorithm 3 is to iteratively increase the user weights such that rate requirements are fulfilled in each step:

---
**Algorithm 3** Minimum Rates Algorithm
---
(1) check feasibility with Algorithm 2
(2) initialize $\boldsymbol{\mu}^* = \boldsymbol{\mu}$
**while** desired accuracy not reached **do**
   **for** $m = 1$ to $M$ **do**
     **if** $R_m < \bar{R}_m$ **then**
       (3) increase $\mu_m^*$ such that $R_m = \bar{R}_m$ using Algorithm 1
     **end if**
   **end for**
**end while**

---

*Theorem 2:* Algorithm 3 converges to a stationary point $\mathbf{R}^*$, which is the global optimum of Problem 3.

The proof is based on the following lemma, which can be verified considering the expression for the rate $R_m$ (21). It is worth noting that the situation is different from that in [20], since the Lagrangian factor $\tilde{\lambda}$ has to be adjusted due to the fixed sum power budget.

*Lemma 6:* For all $m$ and a fixed sum power $\bar{P}$, if the $m$th component $\mu_m$ of the Lagrangian vector is increased and the other components are held fixed, the rate $R_m(\mu)$ remains the same or increases while $R_n(\mu)$ decreases for $n \neq m$.

*Proof:* The rates $R_{n,k}$ (21) depend only on the size of the corresponding sets

$$\mathcal{A}_{n,k} \equiv \{z \in [0, \infty) : u_{n,k}(z) = \max_{i \in \mathcal{M}} u_{i,k}(z)\}. \tag{55}$$

Obviously, an increase of $\mu_m$ does not influence the sets or causes them to shrink for all $n \neq m$ and for all $k$. It remains to show, that an increase of $\mu_m$ can not reduce the set size of the same user $m$ on any subcarrier. Distinguish two cases: if $m \neq \arg\max_{n \in \mathcal{M}} z_0(n) : u_{n,k}(z_0) = 0$, i.e. the marginal utility function of user $m$ does not constitute the last intersection on any subcarrier $k$, the parameter $\tilde{\lambda}$ remains unchanged and hence the sets $\mathcal{A}_{m,k}$ can not shrink. If $m = \arg\max_{n \in \mathcal{M}} z_0(n) : u_{n,k}(z_0) = 0$ on at least one subcarrier $k$, then $\tilde{\lambda}$ increases to meet the sum power constraint. This reduces the maximum intersection $z_0(n,k)$ on all other subcarriers. On the other hand, the intersection of the marginal utility function $u_{m,k}$ of user $m$ is shifted by any $\Delta \mu_m$ and $\Delta \lambda$ about the same offset

$$\Delta z = \frac{\lambda \Delta \mu_m + \mu_m \Delta \lambda}{(\Delta \lambda + \lambda)\lambda} \tag{56}$$

independent of the carrier $k$. The important observation is that this shift is thus equal for all carriers and nonnegative, since the power constraint has to be met and all other carriers experience a power decrease. Hence, the rate of user $m$ can not decrease. ∎

Now, we can proof Theorem 2:

*Proof:* [Theorem 2] Assume that Problem 3 is feasible. Hence, there exist $\lambda \in \mathbb{R}_+$, $\boldsymbol{\mu} \in \mathbb{R}_+^M$ $\tilde{\boldsymbol{\mu}} \in \mathbb{R}_+^M$ such that the KKT conditions are fulfilled. Let $\boldsymbol{\mu}^* = \boldsymbol{\mu} + \tilde{\boldsymbol{\mu}}$. From (21) it can be seen, that $R_m(\mu_m^*)$ is a monotone function in $\mu_m^*$ for any fixed sum power $\bar{P}$, where the power price $\tilde{\lambda}$ is chosen such that the sum power constraint is met. Further,

$$\lim_{\mu_m^* \to \infty} R_m(\mu_m^*) = R_m^{su} \tag{57}$$

where $R_m^{su}$ is the single user water-filling solution for user $m$ and $R_m^{su} \geq \bar{R}_m$ since the problem is feasible by definition. Then in iteration $n+1$ one can find a $\boldsymbol{\mu}^{*(n+1)}$ such that $R_m^{(n+1)} = \bar{R}_m$ in case that $R_m^{(n)} < \bar{R}_m$. Using Lemma 6, $\boldsymbol{\mu}^{*(n)}$ is a component-wise monotone sequence. Define a mapping $U$ representing the update of the sequence $\boldsymbol{\mu}^{*(n)}$. Lemma 6 guarantees that $U$ is order preserving. Starting with $\boldsymbol{\mu}^{*(0)} = \boldsymbol{\mu}$ we know that $U^n(\boldsymbol{\mu}^{*(0)}) \leq \boldsymbol{\mu}_{opt}^*$ for any $n$, where $\boldsymbol{\mu}_{opt}^* \geq \boldsymbol{\mu}$ is the solution such that the KKT conditions are fulfilled and $\mathbf{a} \geq \mathbf{b}$ refers to component-wise greater or equal. Thus, $\boldsymbol{\mu}_{opt}^*$ is a fixed point of the mapping $U$. Hence, $\{\boldsymbol{\mu}^{*(n)}\}$ is a component-wise monotone sequence bounded from above by $\boldsymbol{\mu}_{opt}^*$ and converges to the limiting fixed point $\boldsymbol{\mu}_{opt}^*$ such that the KKT conditions are fulfilled and the associated rate vector $\mathbf{R}^*(\boldsymbol{\mu}_{opt}^*)$ is achieved. ∎

Figures 3 and 4 illustrate the convergence process for an exemplary random channel with $K = 256$ subcarriers and $M = 4$ users. Note, that the algorithm has the same disadvantage as the sum power minimization algorithm based on the BC perspective described in Section IV: For any finite $n$, the resulting point $\mathbf{R}^{(n)}$ is infeasible. Each iteration consists of a loop over all users. Nevertheless, the algorithm offers a considerable speed up compared to the alternative Algorithm 4 based on multilevel rate water-filling presented below. This algorithm is based on eq. (14). Depending on whether the rate constraints are active or not, the water-filling level is adjusted such that the requirements are met.

---

**Algorithm 4** Water-filling based Minimum Rates Algorithm
---
   **(1)** check feasibility with Algorithm 2
   **(2)** choose initial Lagrangian factors $\tilde{\lambda}_+$ and $\tilde{\lambda}_-$
   **while** sum power constraint $\bar{P}$ is not met **do**
      **while** desired accuracy not reached **do**
         **for** $m = 1$ to $M$ **do**
            **(3)** compute the coefficients $n_{m,k}$ (48) for user $m$
            **(4)** do water-filling with fixed level $\mu_m$ (49)
            **if** $R_m < \bar{R}_m$ **then**
               **(5)** choose water-filling level $\log(\mu_m + \tilde{\mu}_m)$ such that $R_m = \bar{R}_m$
            **end if**
         **end for**
      **end while**
      **(6)** increase (decrease) $\tilde{\lambda}$ if $P > \bar{P}$ ($P < \bar{P}$) by bisection
   **end while**
---

We have the following theorem ensuring convergence:

*Theorem 3:* Algorithm 4 converges to a stationary point $\mathbf{R}^*$, which is the global optimum of Problem 3.

*Proof:* First we show that in case of feasibility for any fixed $\tilde{\lambda}$ the inner loop converges: for any fixed $\tilde{\lambda}$, the set of rates fulfulling the rate constraints $\mathcal{R}$ as defined in (50) is a compact convex set. The objective function $g(\mathbf{R}) = \boldsymbol{\mu}^T \mathbf{R}$ is continuously differentiable on $\mathcal{R}$. In each partial step, the unique solution to the convex optimization problem of one specific user $m$ is found. Thus, the same argument as in Theorem 1 holds ensuring convergence to the global optimum $\hat{\mathbf{R}}(\tilde{\lambda})$ due to convexity of the problem.

For the outer loop to converge it remains to show that $P(\tilde{\lambda})$ is monotone in $\tilde{\lambda}$. This can be seen as follows: for each $\tilde{\lambda}$ we have to solve

$$f\left(\tilde{\lambda}\right) = \max_P \left\{\boldsymbol{\mu}^T \mathbf{R}(P) - \tilde{\lambda} P\right\} \tag{58}$$

where $\mathbf{R}(P)$ is the solution to Problem 3 for power budget $P$. Eqn. (58) can be lower bounded by

$$f\left(\tilde{\lambda}\right) \geq \max_P \left\{C_0 \mathbf{R}_0(P) - \tilde{\lambda} P\right\} = f_1\left(\tilde{\lambda}\right)$$

where $\mathbf{R}_0(P)$ is any single user solution and $C_0$ is a constant depending on the weights. Since $f_1(\tilde{\lambda}) \to \infty$ for $\tilde{\lambda} \to 0$ it follows $f(\tilde{\lambda}) \to \infty$ for $\tilde{\lambda} \to 0$ where it is clear that this can only happen if $P \to \infty$. Thus $P \to \infty$ as $\tilde{\lambda} \to 0$ and since the rate and power allocation is unique, $\tilde{\lambda}$ is unique for given $P$ and a smooth function; hence, the claim follows.

Consequently, since the inner loop converges for any fixed value of $\tilde{\lambda}$ and the sum power $P$ is monotone in $\tilde{\lambda}$ and thus can be found by simple bisection, Algorithm 4 converges to a stationary point $\mathbf{R}^*$ being the global optimum. ∎

## C. Lagrangian interpretation and optimal decoding order

Define the vector $\boldsymbol{\mu}^* = (\mu_1 + \tilde{\mu}_1, ..., \mu_M + \tilde{\mu}_M)^T$. The components $\mu_m$ are the weight factors from (11) and the components $\tilde{\mu}_m$ are the Lagrangian multipliers of the rate constraints in (11). Further $\lambda$ is the Lagrangian multiplier due to the power constraint $\bar{P}$. Then the vector $(\boldsymbol{\mu}^*, \lambda)^T$ is the normal vector of the supporting hyperplane to the region $\mathcal{G}(\mathbf{h})$ and the supported point is the solution $(\mathbf{R}^*, \bar{P})$ to (11). The Lagrangian factor $\tilde{\mu}_m$ is strictly greater than zero if and only if the rate constraint of user $m$ is active. Then $\tilde{\mu}_m$ delivers a revenue to the corresponding normal vector of the supporting hyperplane, assuring the minimum rate. This becomes apparent in Figures 5 and 6, where at $\sim 12dB$ the constraint of the first user with $\mu_1 = 0.35$ becomes active. As the SNR approaches the minimum sum power, all existing constraints $\bar{\mathbf{R}}$ become active and the Lagrangian multiplies grow without bounds.

To derive the optimal decoding order, we focus on the Lagrangian interpretation: we know that the vector $\mathbf{R}^*$ is the solution to the optimization problem:

$$\max \quad \boldsymbol{\mu}^{*T} \mathbf{R} \quad \text{subj. to} \quad \mathbf{R} \in \mathcal{C}(\mathbf{h}, \bar{P}) \tag{59}$$

The point is achieved with a decoding order $\pi^*$ such that

$$\mu^*_{\pi^*(M)} \geq \mu^*_{\pi^*(M-1)} \geq \cdots \geq \mu^*_{\pi^*(1)} \tag{60}$$

and the solution $\mathbf{R}^*$ is the vertex $\mathbf{R}^{\pi^*}$ of the corresponding polymatroid. Note, that the optimal decoding order can *not* be determined a priori, since the required weight vector $\boldsymbol{\mu}^*$ — including the Lagrangian factors $\tilde{\boldsymbol{\mu}}$ corresponding to the rate requirements — is not known. The weights $\boldsymbol{\mu}$ constitute only a part of the overall normal vector. Hence, the optimal decoding order is part of the solution. This interesting fact is in analogy to [1], and is illustrated in Figure 5. Note, that there are even for an example with $M = 4$ users five different optimal decoding orders depending on the SNR.

## VII. A UNIFYING FRAMEWORK

It is apparent, that all problems can be solved by algorithms based on iterative multilevel rate water-filling, avoiding convex optimization: In Problem 1, the water-filling levels are known in advance and we perform *multi-level rate water-filling with fixed levels*. In Problem 2, the rate requirements lead to classical *multi-level water-filling*. Finally Problem 3 is solved by combining these two techniques: fixed-level water-filling with an additional optional step ensuring the minimum rate (and hence not to leave the feasible set) with traditional water-filling. This interesting fact is due to the joint problem formulation in eq. (14) on the enhanced set $\mathcal{G}(\mathbf{h})$. Its KKT-conditions characterize the solution to all problems.

A slight disadvantage of the algorithms for Problems 1 and 3 is, that the adequate Lagrangian factor reflecting the sum power constraint is not known: hence, for each step of bisection, e.g. each choice of $\tilde{\lambda}$, the iterative water-filling algorithm has to be run. This is due to the fact, that the reformulation in terms of the rates does not allow an explicit incorporation of the sum power constraint. The minimum sum power algorithm is not affected by this fact, since the Lagrangian factor $\lambda$ is fixed a priori.

On the other hand the downlink formulation from [20] delivers a different class of algorithms: They exploit the marginal utility or rate splitting formulation. The resulting algorithms for Problem 2 and 3 have the slight disadvantage, that for any finite number of steps, the corresponding rate/power allocation does not meet the constraints. Further, the adjustment of the Lagrangian vector $\tilde{\boldsymbol{\mu}}$ requires the calculation of the intersections of the marginal utility functions on all subcarriers in each step.

## VIII. CONCLUSION

In this paper, we considered two fundamental resource allocation problems for OFDM multiuser systems: the sum power minimization problem for required rates and the maximization of a weighted rate-sum for a given power budget. In principle, the OFDM BC is a *non-degraded* channel. We derived the solution to the minimum sum power problem by expressing the dual OFDM MAC channel as a union over the sum of *degraded* channels. This allows to formulate the problem as a convex optimization problem. We showed the coherence of these two problems by interpreting both problems as special cases of an optimization problem in a unifying Lagrangian framework. To this end, we introduced the *enhanced* set $\mathcal{G}(\mathbf{h})$. This framework allows to design alogrithms based on rate water-filling for both problems. In principle, all problems can alternatively be solved in the BC directly using the notion of marginal utility functions stemming from [20]. With these insights, we addressed the *mixed* problem of maximizing a weighted sum of rates under given rate requirements and for a given power budget. Once again, we derived an algorithm from each perspective and showed that the sum of weights $\boldsymbol{\mu}$ and Lagrangian multipliers $\tilde{\boldsymbol{\mu}}$ expanded by the power price $\tilde{\lambda}$ constitutes the normal vector $(\boldsymbol{\mu} + \tilde{\boldsymbol{\mu}}, \tilde{\lambda})^T$ to a tangent hyperplane in $\mathcal{G}(\mathbf{h})$ at the solution. Moreover, the ordering of the Lagrangian multipliers reveals the optimal encoding order for dirty paper coding.

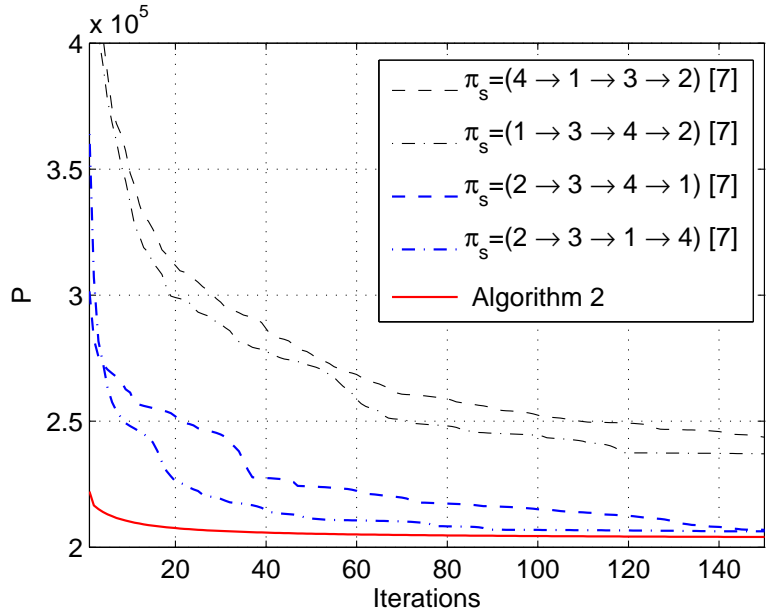

Fig. 1. Comparison of Algorithm 2 with the algorithm for fixed decoding order $\pi_s$ presented in [7]. Random channel with $K = 128$ subcarriers, $M = 4$ users and $\bar{\mathbf{R}} = (2.5\ 0.4\ 0.8\ 2)$ [bps/Hz].

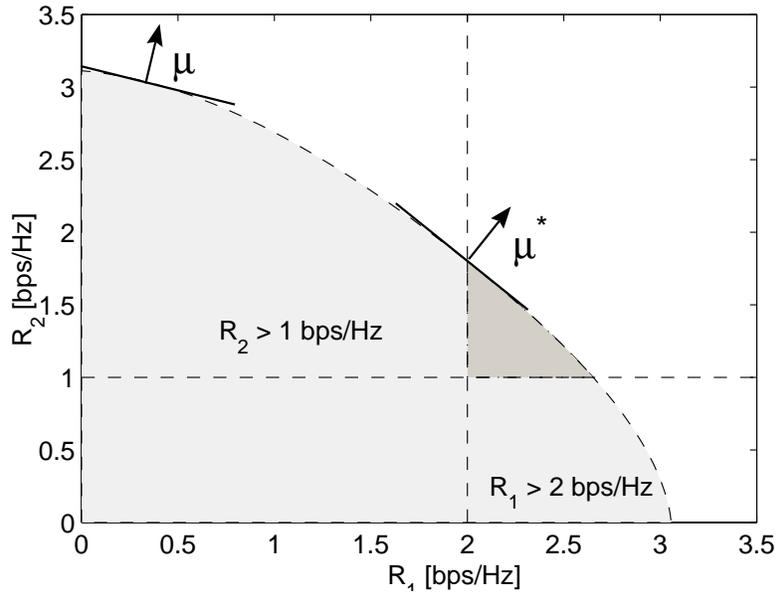

Fig. 2. Feasible set of a random channel with 256 subcarriers and rate requirements $R_1 \geq 2$ and $R_2 \geq 1$ at $10dB$. The vectors $\boldsymbol{\mu}$ and $\boldsymbol{\mu}^*$ illustrate the intial weight vector and the resulting Lagrangian vector $\boldsymbol{\mu}^* = \boldsymbol{\mu} + \tilde{\boldsymbol{\mu}}$.

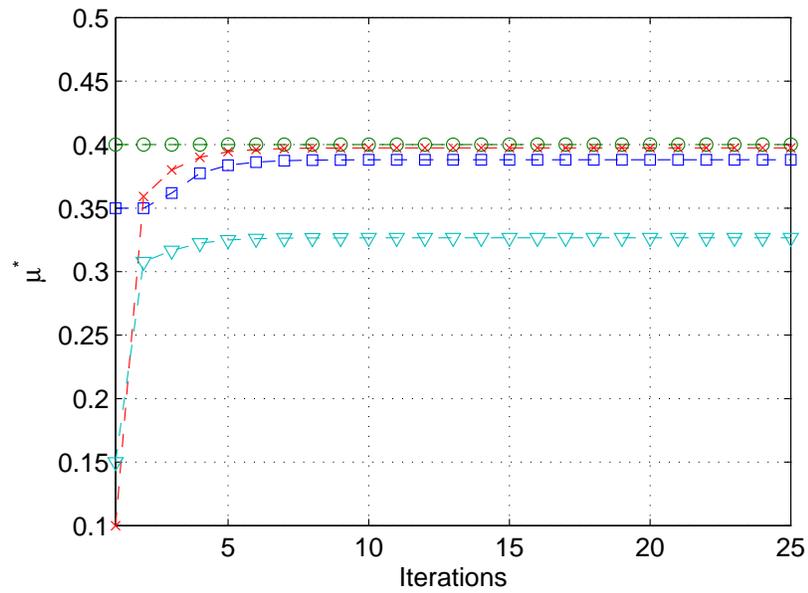

Fig. 3. Convergence of the Lagrangian $\boldsymbol{\mu}^*$ for a system with $M = 4$ users, $K = 256$ at $10dB$ with $\mu = (0.35\ 0.4\ 0.1\ 0.15)^T$ and required rates $\bar{R} = (1\ 0\ 1.25\ 0.5)^T$.

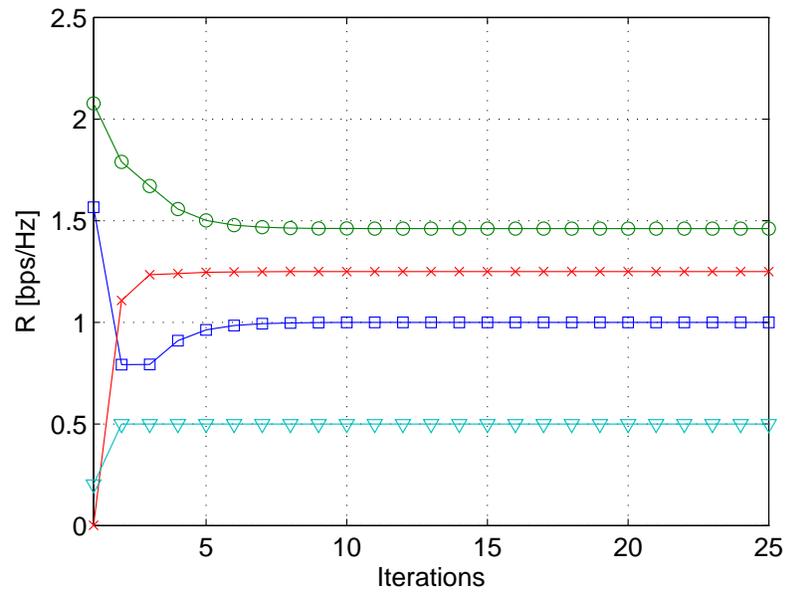

Fig. 4. Convergence of $\mathbf{R}$ for the same system as in Figure 3.

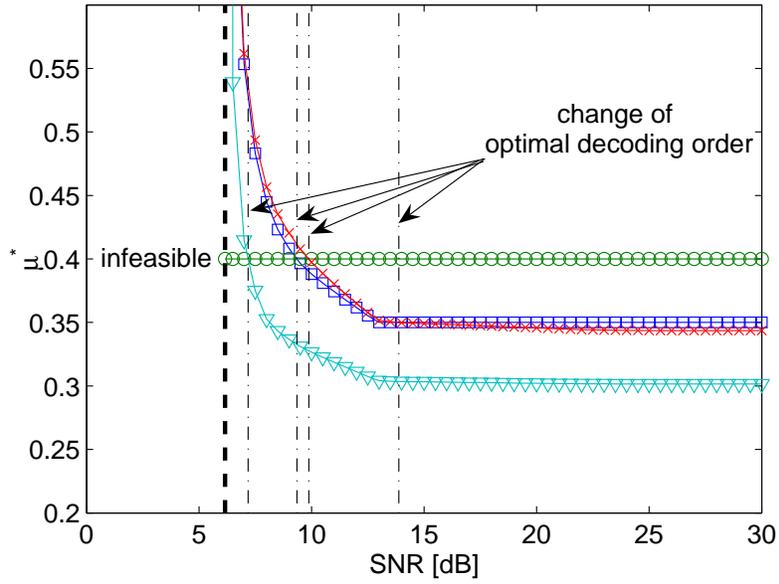

Fig. 5.  $\boldsymbol{\mu}^* = (\boldsymbol{\mu} + \tilde{\boldsymbol{\mu}})$ for different $SNR$ and the same settings as in Figs. 3 and 4. The four SNR-values, where the optimal decoding order changes, are marked with dash-dotted lines.

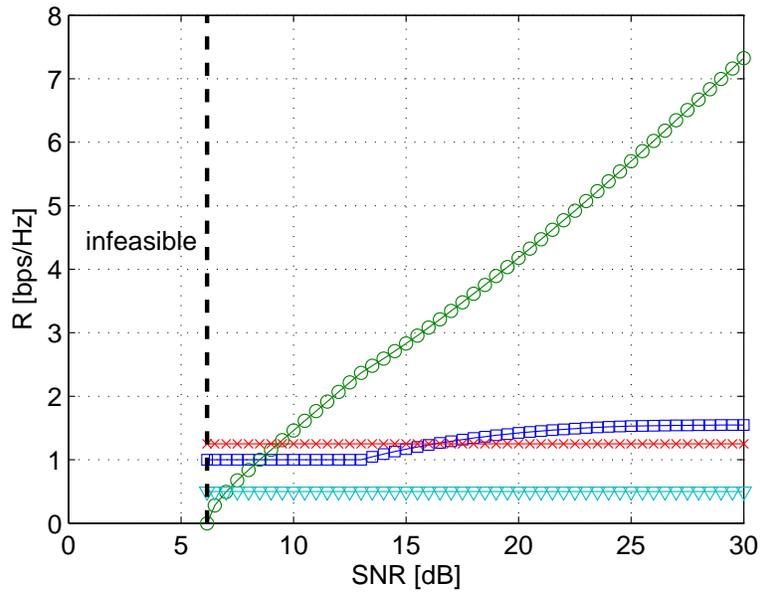

Fig. 6.  Rates $\mathbf{R}$ for different $SNR$ and the same settings as in Figs. 3, 4 and 5.